\documentclass[11pt,onecolumn,conference]{IEEEtran}

\usepackage{graphicx}
\usepackage{vaucanson-g}
\usepackage{float}
\usepackage{citesort}
\usepackage[boxed,linesnumbered,lined]{algorithm2e}

\usepackage{geometry}
\newcommand{\N}{\mathbb{N}}

\newcommand{\paren}[1]{\left( #1 \right)}

\newcommand{\set}[1]{\left\{ #1 \right\}}

\newcommand{\beq}{\begin{eqnarray*}}
\newcommand{\eeq}{\end{eqnarray*}}
\newcommand{\beqn}{\begin{eqnarray}}
\newcommand{\eeqn}{\end{eqnarray}}
\newcommand{\ben}{\begin{enumerate}}
\newcommand{\een}{\end{enumerate}}
\newcommand{\bit}{\begin{itemize}}
\newcommand{\eit}{\end{itemize}}
\newcommand{\hide}[1]{}

\newcommand{\inv}{^{-1}} %

\newcommand{\luniq}{L_{\textrm{{\tiny \textup{UNIQ}}}}}
\newcommand{\lobst}{L_{\textrm{{\tiny \textup{OBST}}}}}

\newcommand{\delim}{\text{{\sf \$}}}
\newcommand{\sigx}[1]{\Sigma_{\bar #1}}

\renewcommand{\th}{\ensuremath{^{\mathrm{th}}}~}

\newcommand{\dpath}{\Rightarrow}

\newcommand{\lsquiggle}{\ensuremath{\stackrel{l}{\rightsquigarrow}}} 
\newcommand{\nsquiggle}[1]{\ensuremath{\stackrel{#1}{\rightsquigarrow}}} 
\newcommand{\charPoly}[2]{\raisebox{0.15em}{$\chi$}_{\mbox{\footnotesize $#1$}}(#2)} 

\usepackage{amssymb}
\usepackage{amsthm}
\usepackage{latexsym}
\usepackage{amsmath}
\usepackage{amsfonts}

 \theoremstyle{plain}
 \newtheorem{thm}{Theorem}
 \theoremstyle{plain}    
 
 \theoremstyle{plain}    
 
 \theoremstyle{remark}
 
 \theoremstyle{remark}
 \newtheorem*{rem*}{Remark}
\theoremstyle{remark}

\newcommand{\bepf}{\begin{IEEEproof}}
\newcommand{\enpf}{\end{IEEEproof}}

\floatstyle{boxed}
\floatname{protocol}{Protocol}
\newfloat{protocol}{tp}{alg}
\begin{document}

\title{Efficiently decoding strings 
from their shingles}

\author{\IEEEauthorblockN{\textbf{Aryeh (Leonid) Kontorovich}}
\IEEEauthorblockA{Email:karyeh@cs.bgu.ac.il\\Computer Science\\
Ben-Gurion University\\Beer Sheva, Israel}
\and
\IEEEauthorblockN{\textbf{Ari Trachtenberg}}
\IEEEauthorblockA{Email:trachten@bu.edu\\
Electrical \& Computer Engineering\\
Boston University\\
Boston, MA 02215}
}

\maketitle

\begin{abstract}
Determining whether an unordered collection of
overlapping substrings (called shingles) can be uniquely decoded into a 
consistent string is
a problem that lies within the foundation of a broad assortment of disciplines
ranging from networking and information theory through cryptography and even
genetic engineering and linguistics.  We present three perspectives on this
problem: a graph theoretic framework due to Pevzner, an automata theoretic
approach from our previous work, and a new insight that yields a
time-optimal 
streaming algorithm for
determining whether a string of $n$ characters over the alphabet $\Sigma$ can
be uniquely decoded from its two-character shingles.
Our algorithm achieves an overall time complexity $\Theta(n)$
and space complexity $O(|\Sigma|)$.
As an application,
we demonstrate 
how this algorithm can be extended to larger shingles for efficient string
reconciliation.

\end{abstract}




\section{Introduction}
\label{sec:intro}


The problem of efficiently reconstructing a string from 
a given encoding
is fundamental to a broad range of settings.
In the information theory world, this is related to the
\emph{$\alpha$-edits} or \emph{string reconciliation}
problem~\cite{orlitsky_focs,DBLP:journals/bioinformatics/ChaissonPT04},
wherein two hosts seek to reconcile remote strings that differ in a fixed
number of unknown edits, using a minimum amount of communication. 
A similar problem is faced in cryptography through fuzzy
extractors~\cite{DBLP:journals/siamcomp/DodisORS08},
which can be used to match noisy biometric data to encrypted
baseline measurements in a secure fashion.  Within a biological
context, this problem has common roots with the sequencing of DNA from
short reads~\cite{DBLP:journals/bioinformatics/ChaissonPT04} and
reconstruction of protein sequences from K-peptides~\cite{shi07}. 
This idea has even shown up in 
computational linguistics, where it was used 
to learn transformations on varying-length sequences~\cite{rumelhart1986}.


In a simple formal statement of the \emph{unique string decoding problem},
one is given a string
$s \in \Sigma^*$ over the alphabet $\Sigma$.  The string is considered
uniquely decodable if there is no other string $s' \in \Sigma^*$
with the same multiset of length $2$ substrings (known as bigrams).  In the general case,
we will be interested in substrings of length $q \geq 2$, which we will
call $q$-grams or \emph{shingles}.  In our analysis, we shall assume throughout
that alphabet characters can be compared in constant time; otherwise, multiplicative
$\log(|\Sigma|)$ terms need to be added where appropriate.
Our main result is a $\Theta(n)$ time, $O(|\Sigma|)$ space streaming algorithm for deciding unique decodability.
To our knowledge, the best previous algorithm \cite{KT12} has time complexity $O(n|\Sigma|^3)$ and space complexity
$\Theta(|\Sigma|^3)$.

\subsection{Approach}
Two principal approaches have been put forth
for deciding unique string decodability.

The first 
is
due to
Pevzner~\cite{springerlink:10.1007/BF01188582} and Ukkonen~\cite{Ukkonen1992191},
who characterized the type of strings that have the same collection of shingles.
This approach can be used to generate a simple unique decodability tester whose
naive worst-case running time on strings of length $n$ is $\Theta(n^4)$.

The second approach is based on an observation that the set of uniquely decodable
strings form a regular language~\cite{DBLP:journals/tcs/Kontorovich04}.  With
this observation, it is possible to produce a deterministic finite state
machine on $\exp(\Omega(|\Sigma|\log|\Sigma|))$ states
\cite{DBLP:journals/jcss/LiX08} 
and a non-deterministic one on
$O(|\Sigma|^3)$ states~\cite{KT12}.
The DFA is prohibitively expensive to construct explicitly, while the NFA
may be simulated in time $O(n|\Sigma|^3)$ and space $\Theta(|\Sigma|^3)$.

In this work, we present a streaming, online, 
linear
time algorithm
for testing unique decodability of a string.  We further show how this
algorithm can be extended to provide an efficient protocol for the
classic $\alpha$-edits (or string reconciliation) problem, in which one is
tasked with reconciling two
remote strings that differ in at most $\alpha$ unknown edits (insertions or
deletions)~\cite{Orlitsky}.  This approach can be extended into a one-way
rateless streaming protocol that 
reconciles strings an arbitrary edit distance apart.

\subsection{Outline}
We begin 
with an overview of related work from the information theory and theoretical
computer science communities
in Section~\ref{sec:related}, 
followed by
a brief exposition
of existing approaches to our core problem
in Section~\ref{sec:existing}.  
Our linear-time algorithm 
for deciding unique decodability,
together with a proof of correctness,
is described in Section~\ref{sec:pseudocode}.
We show in Section~\ref{sec:recon} how this algorithm can be generalized
for the $\alpha$-edits problem,
and close with concluding remarks and remaining open theoretical questions
in Section~\ref{sec:conclusion}.

\section{Related work}
\label{sec:related}

\hide{
We next present some relevant related work.  We begin with an overview
of related work on unique decodability of strings, followed by some
existing results on the one of its applications, the $\alpha$-edits problem.
This leads us to one existing string reconciliation algorithm, whose background
material we cover in this section, but will be more completely covered in Section~\ref{sec:recon}.
}

\subsection{Unique decoding}
It was shown in \cite{DBLP:journals/tcs/Kontorovich04} that 
the collection of strings having a unique reconstruction from the
shingles representation is a regular language.
Following up,
Li and Xie~\cite{DBLP:journals/jcss/LiX08}
gave an
explicit construction of a deterministic finite-state automaton 
(DFA) recognizing 
this language.
Our work in~\cite{KT12} has demonstrated that there is no
DFA of subexponential size for recognizing
this language, and instead we have exhibited an equivalent NFA 
with $\Theta(|\Sigma|^3)$ states.

There has also been work on the probability of a collection of
shingles having a unique reconstruction.  The authors
in~\cite{DBLP:journals/tpds/AgarwalCT06}
show that one can expect a unique decoding for substrings of identically
distributed, independent random bits as long as the substrings are roughly
logarithmic in the size of the overall decoded string. 
The work in~\cite{Dyer1994} also provides evidence of a high probability of
unique decoding for logarithmically sized substrings, and includes generalizations
to non-binary and even non-uniformly random characters for the strings.  This
is extended in~\cite{Arratia200063} to characterize the number of decodings for
a given collection of shingles,
  and 
\cite{Preparata04} considers decoding from
regularly gapped collections of substrings in a DNA sequencing framework.
Finally, 
\cite{AGT12} considers an information-theoretic capacity
of the sequencing problem, and presents a greedy algorithm for reconstruction
that is asymptotically optimal.

\subsection{Edit distance}
The problem of determining the 
minimum
number of edits (insertions or deletions)
required to transform one string into another has a long history in the
literature~\cite{CLRS,DBLP:books/cu/Gusfield1997}.  
Orlitsky~\cite{orlitsky_focs} shows that the
amount of communication $C_{\hat{\alpha}}(x,y)$ \emph{necessary} to reconcile two strings
$x$ and $y$ (of lengths $|x|$ and $|y|$ respectively) that are known to be
at most $\hat{\alpha}$-edits apart is at most
\[
C_{\hat{\alpha}}(x,y) \leq f(y) + 3 \log f(y) + \log \hat{\alpha} + 13,
\]
for
\[
\log\left( \binom{|y|+\hat{\alpha}}{\hat{\alpha}} \right) \leq \lceil f(y) \rceil \leq
\log \left( \binom{|y|+\hat{\alpha}}{\hat{\alpha}} \right) + 3\log(\hat{\alpha}),
\]
although he leaves an efficient one-way protocol as an open question.

The literature includes a variety of proposed protocols for this problem.
Cormode et al.~\cite{DBLP:conf/soda/CormodePSV00} propose a hash-based approach 
that requires a known
bound $\hat{\alpha}$ on edits between $x$ and $y$ (assuming, without loss of
generality, that $y$ is the longer string) and communicates at most
\begin{equation}
\label{eq:cpsv00}
4\alpha \log( \frac{2|y|}{\alpha} ) \log(2 \hat{\alpha}) + O \left(\alpha \log |y| \log \frac{\log(|y|)}{\ln \frac{1}{1-\epsilon}} \right)
\end{equation}
 bits to reconcile the strings with probability of failure $\epsilon$.

Orlitsky and Viswanthan~\cite{orlitskyIsit} propose a interactive
protocol that does not need to know the number of edits in advance and
requires at most
\[
2 \alpha \log |y| \left( \log |y| + \log \log |y| + \log(1/\epsilon) + \log \alpha \right)
\]
bits of communication.

Other approaches include those of Evfimievski~\cite{Evfimievski} for
small edit distances, Suel~\cite{DBLP:conf/icde/SuelNT04} based on
delta-compression, and Tridgell~\cite{rsync} which presents the
computationally efficient (but potentially communicationally inefficient)
rsync protocol.

\subsection{Reconciliation}
\label{subsec:recon}
Another natural approach to the $\alpha$-edits problem involves the utilization of
a \emph{reconciliation} algorithm, which reconciles remote data with minimum communication.

\paragraph{Set reconciliation}
The problem of set reconciliation seeks to reconcile two remote sets $S_A$ and $S_B$ of
$b$-bit integers using minimum communication.  The approach in~\cite{MTZ00} involves
translating the set elements into an equivalent \emph{characteristic polynomial}, so
that the problem of set reconciliation is 
reduced to
an equivalent problem of
rational function interpolation, much like in Reed-Solomon decoding~\cite{MWS}.

The resulting algorithm requires one message
of roughly $b m$ bits of communication and $b m^3$ computation 
time
to reconcile two sets
that differ in $m$ entries.  
The approach can be improved to
expected $b m$ communication and computation through the use of
interaction~\cite{practical_allerton} and generalized to multisets 
and to arbitrary error-correcting codes~\cite{KLT_allerton}.

\paragraph{String reconciliation}
A string $\sigma$ can be transformed into a multiset $S$ through {\em shingling}, or collecting
all contiguous substrings of a given length, including repetitions.  
For example, shingling the string {\sf katana}
into length $2$ shingles produces the multiset:
\begin{equation}
\label{eq:shingles}
\left\{ {\sf at}, {\sf an}, {\sf ka}, {\sf na}, {\sf ta} \right\}.
\end{equation}
As such, in order to reconcile two strings $\sigma_A$ and $\sigma_B$, the protocol
STRING-RECON~\cite{DBLP:journals/tpds/AgarwalCT06}
first shingles each string, then reconciles the resulting sets, and then
puts the shingles back together into strings in order to complete the reconciliation.  It
is important to note that if two strings differ by $\alpha$ edits, then they will also
differ in $O(\alpha)$ shingles, as long as shingle size is a constant.

The process of combining shingles of length $l$ back into a string involves the construction
of a modified de Bruijn graph of the shingles.  In this graph, each shingle corresponds to
an edge, with weight equal to the number times the shingle occurs in the multiset.  The
vertices of the graph are all length $l-1$ substrings over the shingling alphabet; in this
manner, an edge $e(u,v)$ corresponds to a shingle $s$ if $u$ (resp. $v$) is a prefix
(resp. suffix) of $s$.  A special character $\delim$ used at the beginning and end of
the string in order to mark the first and last shingle.

An Eulerian cycle in the modified de Bruijn graph, starting
at the first shingle, necessarily corresponds to a string that is consistent with the
set of shingles.  Unfortunately, there may be a large number of strings consistent
with a given shingling, so that well-defined decoding requires 
either the specification of one cycle of interest or another way to guarantee
only one possible cycle.

\section{Existing approaches}
\label{sec:existing}
We now describe two existing approaches for determining whether a given
string is uniquely decodable.

\subsection{Transformation}
In an analysis of approximate string matching, Ukkonen~\cite{Ukkonen1992191}
conjectured that two strings with the same shingles are related
through two string transformations, for $(q-1)$-grams $z_1$ and $z_2$ and
arbitrary strings $x_i$:

\begin{itemize}
\item \textbf{Transposition} - wherein a string
\[
x = x_1 z_1 \mathbf{x_2} z_2 x_3 z_1 \mathbf{x_4} z_2 x_5
\]
is transformed into
\[
x' = x_1 z_1 \mathbf{x_4} z_2 x_3 z_1 \mathbf{x_2} z_2 x_5.
\]

\item \textbf{Rotation} - wherein a string
\[
x = z_1 x_1 z_2 x_2 z_1
\]
is transformed into
\[
x' = z_2 x_2 z_1 x_1 z_2.
\]
\end{itemize}

Pevzner~\cite{springerlink:10.1007/BF01188582} proved that this conjecture
is true, thus providing a simple but inefficient algorithm for determining
the unique decodability of a string.

\subsection{Regular languages}
\label{subsec:regular}
A second approach for testing unique decodability is automata theoretic in nature.

\subsubsection{Preliminaries}
We assume
a finite alphabet $\Sigma$
along with a special delimiter character $\delim\notin\Sigma$,
and define $\Sigma_\delim=\Sigma\cup\set{\delim}$. 
For $k\ge1$,
the
$q$-{gram map} $\Phi$
takes
string $x\in\delim\Sigma^*\delim$ 
to
a vector $\xi\in\N^{\Sigma_\delim^q}$,  
where $\xi_{i_1,\ldots,i_q} \in \N$ 
is the number of times the string $i_1\ldots i_q\in\Sigma^q$ occurred in
$x$ as a contiguous subsequence, counting overlaps.  Note that, though
we focus this section on the \emph{bigram} case when $q=2$, the results
are straightforwardly generalized to the case $q>2$.

It is easy to see that 
the bigram map $\Phi:\delim\Sigma^*\delim\to\N^{\Sigma_\delim^2}$ is not 
injective; for example, the shingles in~\eqref{eq:shingles} imply that
$\Phi(\delim {\sf katana}\delim)=\Phi(\delim{\sf kanata}\delim)$. 
We denote by 
$\luniq\subseteq\Sigma^*$
the collection 
of all strings $w$ for which
$$\Phi\inv(\Phi(\delim w\delim))=\set{\delim w\delim}$$
and refer to these strings as {\em uniquely decodable}, meaning that there is
exactly one way to reconstruct them from their bigrams. 
The induced  {\em bigram graph} of a string $w\in\Sigma^*$ 
is a weighted directed graph $G=(V,E)$,
with $V=\Sigma_\delim$ and $E=\set{e(a,b):a,b\in\Sigma_\delim}$,
where the edge weight
$e(a,b)\ge0$ records the number of times $a$ occurs immediately before $b$
in the string $\delim w\delim$. 
%
Finally, we will denote the 
omission 
of a symbol 
from the alphabet by
$\sigx{x}:=\Sigma\setminus\set{x}$ for $x\in\Sigma$.
%
%
%
%

%
%
%
%
%

\subsubsection{Regularity of obstructions}

%
%

For $x\in\Sigma$ and $a,b\in\sigx{x}$, the languages
$$I_{x,a,b} = L \left( { \Sigma^* a  x \sigx{a}^* b \Sigma^* }\right)$$
and
$$J_{x,a,b} = L \left( { \Sigma^* a \sigx{x}^* b \Sigma^* } \right)$$
form the obstruction language
$$K_{x,a,b}=I_{x,a,b}\cap J_{x,a,b},$$
whose elements are called {\em obstructions} (because they obstruct a unique
decoding).  
The language of all obstructions is thus 
\beqn
\label{eq:obst}
\lobst = 
\bigcup_{x\in\Sigma} \bigcup_{a,b\in\sigx{x}} K_{x,a,b}.
\eeqn

The work in~\cite{KT12} provides a canonical DFA that recognizes $K_{x,a,b}$ with
$9$ states, 
regardless of $\Sigma$. 
Over all $x\in\Sigma$ and $a,b\in\sigx{x}$,
there are
\beqn
\label{eq:nobstr}
|\Sigma|
\paren{ |\Sigma|-1 + 
(|\Sigma|-1)(|\Sigma|-2)
 }
\eeqn
distinct obstruction languages, whose union can thus be accepted by an NFA of
$\Theta(|\Sigma|^3)$ states.

The main theorem in that work is that the language of obstructions
is precisely the complement of the language of uniquely decodable
strings.
\begin{thm}[\cite{KT12}]
\label{thm:luniqlobst}
$$ \lobst = \Sigma^* \setminus \luniq.$$
\end{thm}

The result of Theorem~\ref{thm:luniqlobst} is that the NFA accepting $K_{x,a,b}$'s can be
used to test for unique decodability.




\section{Efficient online testing}
\label{sec:pseudocode}
We now describe our main result:  an efficient, online streaming algorithm for
determining whether a given string $w\in\Sigma^*$ is uniquely decodable
from its bigrams.
Algorithm~\ref{alg:unique} is online in the sense that it needs only
constant-time pre-processing, and streaming, in that results for one string can be
sub-linearly extended to a superstring.

As a convention, we will use ``low'' letters $a,b,c$ to denote members of $\Sigma$ while the 
``high'' letters $u,v,w$ will denote \emph{strings} over $\Sigma$. 
For any $u\in\Sigma^*$, we write $G(u)$ for the bigram graph induced by $u$, and we shall
use the notation $a\to b$ (resp. $a \dpath b$) to mean that there is a directed edge
(resp. path) from $a$ to $b$.
We use the shorthand
``$u$ is UD'' to denote that $u\in\luniq$. The $i$\th character of $w$ is denoted by 
$w[i]$ 
and characters $i$ through $j$ by $w[i:j]$.


\IncMargin{1em}
\begin{algorithm}
\SetKwData{True}{TRUE} \SetKwData{False}{FALSE}
\SetKwData{Yes}{YES} \SetKwData{No}{NO}
\SetKwInOut{Input}{input}\SetKwInOut{Output}{output}
\Input{string $w\in\Sigma^*$}
\Output{\True if $w\in\luniq$ and \False otherwise}
\BlankLine
initialize each $v\in\Sigma$ as {\bf not} having been visited\;
initialize each $v\in\Sigma$ as {\bf not} belonging to a cycle\;
initialize the graph $G$ with vertex set $\Sigma$ and no edges\;
mark $w[1]$ as having been visited\;
\For{$i:=2$ \KwTo $|w|$}{
    has the node $w[i]$ already been visited?
    \eIf{\No}{
        mark $w[i]$ as having been visited\;
    } (\tcp*[h]{node $w[i]$ has already been visited -- thus, it is on a cycle})
    { 
        does the edge $w[i-1]\to w[i]$ already exist in $G$?
        \eIf{\No}{
            does $w[i]$ belong to an existing cycle?
            \eIf{\Yes}{
                \tcp{intrusion on an existing cycle}
                \Return{\False}\;    \label{break-intrusion}
            } (\tcp*[h]{creating a new cycle})
            { 
                label $w[i]$ and all the nodes visited since the previous   \
                occurrence of $w[i]$ as belonging to a cycle\;
            } 
        }
        {
            \tcp{the edge $w[{i-1}]\to w[i]$ already exists in $G$\\
            stepping along an existing cycle}
        }
    }
    are there two distinct nodes $a,b\in G$
        such that $a\to w[i],b\to w[i]$
        and $a,b$ belong to the same strongly connected component of $G$?\\
        \tcp{the possibility $a=w[i]$ is not excluded}
    \If{\Yes}{
        \Return{\False}\;      \label{break-ccparents}
    }
    draw the edge $w[i-1]\to w[i]$ in $G$\;
}
\Return{\True}\;
\caption{Online algorithm for testing unique decodability}
\label{alg:unique}
\end{algorithm}

The following theorem establishes the correctness of Algorithm~\ref{alg:unique}.
\begin{thm}
Algorithm~\ref{alg:unique} returns TRUE iff its string $w \in \luniq$.
\end{thm}

\bepf
Observe that $u\notin\luniq$ implies $u\sigma\notin\luniq$ for all $\sigma\in\Sigma$ (in fact, $\Sigma^*\setminus\luniq$ is 
a two-sided
ideal under concatenation). Thus, as soon as a non-UD prefix is observed, we know that the entire string is not UD.

Our algorithm can conclude that the prefix is not UD in two places: at 
line \ref{break-intrusion} 
and 
line \ref{break-ccparents}. 
Line \ref{break-intrusion}
handles an intrusion upon an existing cycle. Formally, this means that the prefix 
$u=w[1:i-1]$ may be expressed as $u=v a_1 a_2 \ldots a_k v'$ where 
$v,v'\in\Sigma^*$, $(a_j)_{1\le j\le k}\in\Sigma$,
$a_1=a_k$ and the current character
$x=w[i]$ is equal to some $a_j$, for $1\le j\le k$. Thus the string
$w[1:i]$ has at least two distinct decodings, among which are
$v a_1 a_2 \ldots a_k v'x$ and
$v a_1 v' x a_{j+1} a_{j+1}\ldots a_k a_2 \ldots a_j$.

Line \ref{break-ccparents}
handles the case of \emph{communicating} parents $a$ and $b$ (one of them possibly a self-parent), by which we mean
that they are in the same strongly connected component. 
Note that the mere existence
of a node with two communicating parents is insufficient to disqualify a string, as the example
$w=axbxa$ shows. However, the condition in the loop has us visiting a node $x=w[i]$ that {\em already} has
two communicating parents. 
First, let us dispense with the case where $x\in\set{a,b}$ --- say, $x=b$ without loss of generality. Since 
$a\to x$, $x\to x$ and $x\dpath a$, 
the self-loop at $x$ can be taken after the first visit to $x$ or after a later visit, creating an ambiguity in the decoding.
Thus, we will take $x\notin\set{a,b}$ and
assume without loss of generality that 
the first occurrence of $a$ in $w[1:i-1]$ occurs before the first occurrence of $b$.
We claim that $w[1:i]$ must be of the form
$(\Sigma\setminus\set{a,b,x})^* 
a x 
(\Sigma\setminus\set{a,x})^* 
b x
(\Sigma\setminus\set{a})^* 
ax$.
Indeed, $a$ must occur twice in $w[1:i-1]$ (since it occurs before $b$ but $b\dpath a$)
and the second occurrence of $a$ must be after the last occurrence of $b$
(for otherwise $b$'s directed path to $a$ would intrude upon an existing cycle and
render the string non-UD earlier on).
Immediately following $a$'s first occurrence is $x$, followed by some string 
containing neither $a$ (whose second occurrence will be at $w[{i-1}]$)
nor $x$ (for otherwise the edge $b\to x$ will intrude on an existing cycle and disqualify the string
earlier on). 
Then $bx$ occurs for the first time and is immediately followed 
a string not containing $a$ and followed by $ax$.
Define $i_1,i_2,i_3$ to be the indices of the first, second and last occurrences of $x$ in
$w[1:i]$, respectively, and put 
$v_1=w[1:i_1-1]$, 
$v_2=w[i_1+1:i_2-1]$,
$v_3=w[i_2+1:i_3-1]$.
Observe that $w[1:i]=v_1 x v_2 x v_3 x$ 
and $w'\equiv v_1 x v_3 x v_2 x$ have the same bigram encoding.
Note also that necessarily $v_2\neq v_3$, since the former does not contain $a$ and the latter does.
This shows that the prefix $w[1:i]$ is not UD.

Having shown that whenever our algorithm disqualifies a string it is indeed not UD (\emph{completeness}), 
we now show that any 
string that survives at the loop's termination is in fact UD (\emph{soundness}).

We prove this claim by induction on the prefix length.
Our inductive hypothesis is that the prefix $u=w[1:i-1]$ is UD. 
We read the next character $x=w[i]$. 
Clearly, if $u\in\luniq$ and $x$ does not occur in $u$ then $ux\in\luniq$.
As such, we consider the case where $x$ does occur somewhere in $u$. 
If the edge $w[i-1]\to x$ does not currently exist in the bigram graph $G(u)$,
then we may assume that $x$ occurs exactly once in $u$,
as, otherwise,
it would already be marked as belonging to a cycle, disqualifying $u$.
Thus, $u=vxv'$ where $v,v'\in\sigx{x}^*$. 
Furthermore, our assumption that
$u\in\luniq$ implies that $v$ and $v'$ cannot have any letters in common,
for then $x$ would be on a cycle 
upon which the new edge $w[i-1]\to x$ would intrude.
Finally, observe that if two UD strings have no characters in common, then their concatenation is also UD.
Thus, $ux\in\luniq$.

It remains to consider the case where the edge 
$w[i-1]\to x$ already exists in $G(u)$. Although we are stepping along an existing cycle
and not creating a new one,
this transition may render the string non-UD, as the example
$w[1:i]=axbxbax$ shows.
Since $u=w[1:i-1]$ is UD, there can be at most two distinct $a,b$ such that $a\to x$ and $b\to x$
(the existence of 3 or more distinct nodes pointing to $x$ is easily seen to render $u$ non-UD; see
\cite[Theorem 9]{DBLP:journals/tcs/Kontorovich04} for an analogous fact regarding 3 or more children).
The case of a single $a\to x$ is trivial, so suppose that
$a\to x$ and $b\to x$, but $a$ and $b$ are not in the same strongly connected component.
There is no loss of generality in assuming that $b$ is reachable from $a$ but not vice versa.
In this case, the only valid decoding of $w[1:i]$ is of the form
$vaxv'bx$ where 
$v\in(\Sigma\setminus\set{x,b})^*$ and 
$v'\in(\Sigma\setminus\set{a})^*$.
\enpf

\subsection{Runtime analysis}
\label{subsec:runtime}
Algorithm~\ref{alg:unique} can be implemented in time $\Theta(n)$ on
strings of $n$ characters over an alphabet $\Sigma$,
with the aid of several simple data structures.  We
account for the running time:

\begin{itemize}
\item Lines 01-04.  This is simple initialization.  It can be accomplished explicitly in $\Theta(|\Sigma|)$ time for our data structures delineated hereafter, or in constant time with a sparse representation.
\item Lines 06-08.  We use a simple array to keep track of which vertices have been seen, a constant time cost for each string character.
\item Line 9.  The key observation here is that the graph is necessarily sparse, since any node with more
than two parents or children necessarily renders the graph not uniquely decodable~\cite{DBLP:journals/tcs/Kontorovich04}.  As such, the graph can be stored as an adjacency list so that this line represents a constant time operation for each string character.
\item Lines 10-19.  We maintain a stack onto which vertices are pushed in the order that they are visited.  When a vertex is visited a second time, we pop all vertices off the stack until we revisit the original node, marking all popped vertices as being within an existing cycle.
Each character of $w$ will be, at worst, pushed and popped from the stack once, resulting in an aggregated running time of $\Theta(n)$ for this step.
\item Line 21.  To determine whether two vertices 
$a,b$
are in the same strongly connected component, we record the first 
and last index in $w$ at which $a$ occurs in $i_a$ and $j_a$, respectively,
and do the same for $b$.
The vertices $a$ and $b$
belong to
the same connected component
if and only if
$[i_a,j_a]\cap[i_b,j_b]\neq\emptyset$.
This check is a constant-time operation per character.
\end{itemize}

\section{String reconciliation}
\label{sec:recon}
We next present the string reconciliation protocol in~\cite{ISIT2012} as a specific
example where our online unique decodability algorithm is applicable.  This specific
protocol is 
a refinement of a shingling approach in~\cite{DBLP:journals/tpds/AgarwalCT06},
and is further based on a transformation to an instance of the set reconciliation~\cite{MTZ00}.


\subsection{Definitions}
The protocol is fundamentally based on the concept of 
a \emph{shingling}.  Formally, 
a {\em shingle} $s=s_1s_2\ldots s_k$ is simply an element of $\Sigma_\delim^*$.
For two shingles
$s=s_1s_2\ldots s_k$ and $t=t_1t_2\ldots t_\ell$, we write
$s  \lsquiggle t$ if there is some length $\geq l$ suffix $u$ of $s$ that is
also a prefix of $t$, or, more precisely, if we can rewrite
$s = s' u$ and $t = u t'$ for  strings $s', t'$ and $|u| \geq l$.
We define the {\em non-overlapping concatenation} $s\odot_l t$ (or just
$s \odot t$ in context) as the
concatenation $s'ut'$,
where $s=s'u$, $t=ut'$ and $|u|=l-1$.
For example, ${\sf kata}\nsquiggle{3}{\sf tana}$
and ${\sf kata}\odot_3 {\sf tana}={\sf katana}$.

For a fixed $l$, the sequence of shingles 
$s^1 \lsquiggle s^2 \lsquiggle \ldots \lsquiggle s^t$
is said to {\em represent} the word $w\in\Sigma^*$ if 
$w=s^1 \odot s^2 \odot \ldots \odot s^t$ and $s^i \lsquiggle s^{i+1}$ for all $i$. 
If $S=\set{s^1,\ldots,s^t}$ is a multiset of shingles, we will 
use $\Phi\inv(S) \subseteq \Sigma^*$ to denote
the collection of all words represented by $S$.
More formally, define $\Pi=\Pi(S)$ to be the set of all permutations on $t=|S|$ elements
with the property that $s^{\pi(i)} \lsquiggle s^{\pi(i+1)}$ for all $i$. 
Then $\Phi\inv(S)$ is
$$ 
\set{w\in\Sigma^* : \delim w\delim=
s^{\pi(1)} \odot 
s^{\pi(2)} \odot 
\ldots 
\odot 
s^{\pi(t)},
\pi\in\Pi}.
$$
We refer to the members of $\Phi\inv(S)$ as the {\em decodings} of $S$, and say that $S$ is
uniquely decodable if $|\Phi\inv(S)|=1$.

A {\em shingling} $I$ of a word $w=w_1\ldots w_t\in\Sigma^*$ is a set of
substrings of $w$ that represents $w$.  
We say that $I$ is an uniquely decodable shingling of $w$ if $|\Phi\inv(I(w))|=1$.

As a simple example, consider the string $w={\sf katana}$ with the shingling 
$I(w) = \set{
\delim{\sf k},
{\sf ka},
{\sf at},
{\sf ta},
{\sf an},
{\sf na},
{\sf n}\delim}$.
As we saw in Section~\ref{subsec:regular}, for $l$=2, $I$ can be alternately decoded
into ${\sf kanata}$ and is thus not uniquely decodable.  However, if the second and
third shingles are merged into ${\sf ata}$, that the shingling becomes
$\set{
\delim{\sf k},
{\sf ka},
{\sf ata},
{\sf an},
{\sf na},
{\sf n}\delim}$,
and then there is exactly one decoding: ${\sf katana}$.




\begin{protocol}
\caption{Reconciliation of remote strings $\sigma$ and $\tau$.\label{protocol:string}}
\begin{itemize}
\item[1.]  Split $\sigma$ into a set $S_\sigma$ of length $l$ shingles, with the $i$\th shingle of the string denoted $s_i$.  Similarly split $\tau$ into $S_\tau$.
\item[2.]  Reconcile sets $S_\sigma$ and $S_\tau$.
\item[3.]  The first host sets $S^0_\sigma \longleftarrow \{ s_0 \}$.
\item[4.]  \textbf{For} $i$ from 1 to $|\sigma|-l+1$ \textbf{do}
\begin{itemize}
\item[]  $S^i_\sigma \longleftarrow S^{i-1}_\sigma \cup \{ s_i\}$
\item[]  \textbf{While} $S^i_\sigma$ is not uniquely decodable
\begin{itemize}
\item[] Merge the last two shingles added to $S^i_\sigma$.
\end{itemize}
\end{itemize}
\item[5.]  Exchange indices of merged shingles.
\item[6.]  Uniquely decode $S^i_\sigma$  and $S^i_\tau$ on the remote hosts.
\end{itemize}
\end{protocol}

\subsection{Elaboration}
Protocol~\ref{protocol:string} transforms a string that is not uniquely decodable
into one that is by merging shingles.  
Several important details of Protocol~\ref{protocol:string} require explanation
and proof of correctness.


\subsubsection{Steps 1 and 2}
The first two steps of the protocol derive from the base protocol described in Section~\ref{subsec:recon}.
Note that $l$ is an implementation parameter.

\subsubsection{Step 3}
The expression $S^i_\sigma$ represents the multiset of shingles that have been seen so far. 
It is modified, by combining shingles as necessary in the subsequent steps, in order to
ensure unique decodability.  If full reconciliation is desired (i.e. both hosts know the other host's
string, as opposed to just one host having this knowledge) then Steps 3 and 4 are similarly
run on the remote host with set $S^i_\tau$.

\subsubsection{Step 4}
In merging two shingles $s_a$ and $s_b$, we are simply computing the non-overlapping
concatenation $s_a := s_a \odot s_b$, as defined earlier.  Since the shingles are contiguous
and based on an initial length $l$ shingling, we know necessarily that $s_a \lsquiggle s_b$. 
Furthermore, it is clear that such merging will always, eventually,
lead to a decodable set of shingles because, at worst, the protocol results in just one shingle
representing the entire string, which is necessarily uniquely decodable.

The main challenge of this step is in checking whether a given set of shingles is uniquely
decodable.  This can be done in an online manner with two extensions to our algorithm in~\ref{sec:pseudocode}.

\paragraph{Extension to $q$-grams}
First, Algorithm~\ref{alg:unique} needs to be extended to shingles of length
$q>2$, rather than just bigrams.  This can be accomplished by considering $w_i$
to be the length $q-1$ prefix of the $i$\th shingle of the input string;
for $q=2$, we have the existing case that $w_i$ is the $i$\th character of the string.

In this model, the input alphabet is enlarged to $\Sigma^{q-1}$ and edges correspond
to shingles.  Note that this extension works even with the mixed-length shingles
which Protocol~\ref{protocol:string} produces.

\begin{figure}[t]
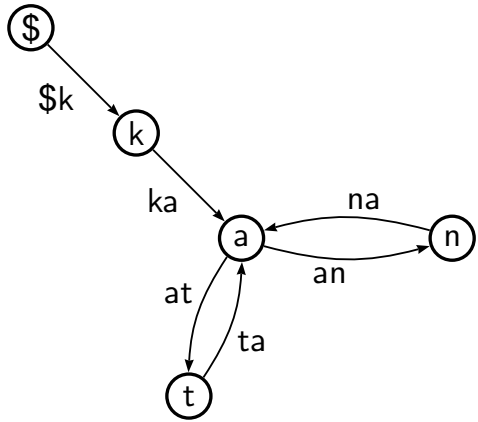

\centering
\scalebox{0.7}{
\begin{VCPicture}{(-6,-4)(5,4)}
\State[\delim]{(-5,4)}{0}
\State[{\sf k}]{(-3,2)}{k}
\State[{\sf a}]{(-1,0)}{a}
\State[{\sf t}]{(-2,-3)}{t}
\State[{\sf n}]{(3,0)}{n}
\EdgeR{0}{k}{\delim {\sf k}}
\EdgeR{k}{a}{{\sf ka}}
\ArcR{a}{t}{{\sf at}}
\ArcR{t}{a}{{\sf ta}}
\ArcR{a}{n}{{\sf an}}
\ArcR{n}{a}{{\sf na}}
\end{VCPicture}
}
\caption{A de Bruijn graph corresponding to the substring ${\sf \delim katana}$.}
\label{fig:nonUD}
\end{figure}

\begin{figure}[t]
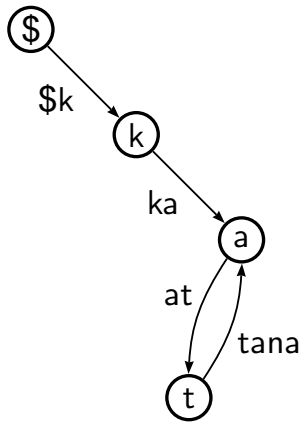

\centering
\scalebox{0.7}{
\begin{VCPicture}{(-6,-4)(5,4)}
\State[\delim]{(-5,4)}{0}
\State[{\sf k}]{(-3,2)}{k}
\State[{\sf a}]{(-1,0)}{a}
\State[{\sf t}]{(-2,-3)}{t}
\EdgeR{0}{k}{\delim {\sf k}}
\EdgeR{k}{a}{{\sf ka}}
\ArcR{a}{t}{{\sf at}}
\ArcR{t}{a}{{\sf tana}}
\end{VCPicture}
}
\caption{A uniquely-decodable modified de Bruijn graph corresponding to the substring {\sf \delim katana}.}
\label{fig:decode}
\end{figure}

\paragraph{Extension for shingle merging}
When shingles are merged, we are effectively combining two edges $e_1=(v_1,v_2)$
and $e_2 = (v_2,v_3)$ into their transitive closure $e_3 = (v_1,v_3)$.  This is
demonstrated when Figure~\ref{fig:nonUD}, which is not uniquely decodable,
is transformed into Figure~\ref{fig:decode}, which is uniquely decodable because
shingles \texttt{ta},\texttt{an}, and \texttt{na} have been merged into their
transitive closure \texttt{tana}.

Such a transitive closure can be implemented in Algorithm~\ref{alg:unique} by
patching steps 11 and 20 so as
to reverse one step of the broader iteration, and add the transitive closure
edge instead, instead of returning \texttt{FALSE} as in the current implementation.


\subsubsection{Step 5}
Each host needs to know which shingles were merged on the other host in order to
produce a uniquely decodable multiset of shingles.  To exchange this information, we
first canonically order all shingles, and then note that each merge involves at
least one shingle of length $l$ and another (possibly composite) shingle of length $\geq l$.  As such, a merge is fully specified by sending the index of the length $l$ shingle,
and the index of one of the shingles that comprises the composite shingle.


\subsubsection{Step 6}
The resulting collection of shingles can only be decoded in only one way, which can be provided
by any efficient algorithm for generating an Eulerian cycle through the graph (e.g., the algorithm implied in
\cite[Theorem 11]{DBLP:journals/tcs/Kontorovich04} can be implemented in linear time).



\subsection{Communication Complexity}
Only Steps  2 and 5 in Protocol~\ref{protocol:string} transmit data.
For two strings of length $n$ differing in $\alpha$ edits, Step 2 will require
$O(\alpha \; l^2)$ bits 
 of communication for the implementation parameter $l$.  Step 5
will require between $0$ and $2n \log (n-l+1)$ communication, depending on the decodability
of the string.

More precisely, the communication efficiency of the protocol relies upon having as few merge
operations as possible, since,
at worst, \emph{every} shingle is merged in Step 5, requiring $2n \log n$ bits of communication
for a shingle set of size $n$.  In the best case, no shingles are merged and the
communication complexity of the protocol is directly related to the edit distance
between reconciled strings.  The shingle size $l$ thus represents a tradeoff between
communication spent on set reconciliation and communication spent on merge identification.

Though it is hard to give precise bounds on the number of shingles that are merged
in this step, the work in~\cite{DBLP:journals/tpds/AgarwalCT06} provides some guidance
for random strings.  Specifically, for strings of $n$ random bits, in which each bit is $0$ with
probability $p>0.5$, then we can expect each node in the de Bruijn graph of length $l$ shingles
to have only one outgoing edge (implying unique decodability) if
\begin{equation}
\label{eq:shingleSize}
l \leq n + 1 + \frac{W \left(-\ln(p)p^{-n} \right)}{\ln p},
\end{equation}
where $W(\cdot)$ is the 
Lambert $W$ function
\cite{MR1414285}.  When $n$ goes to infinity,
then~\eqref{eq:shingleSize} is $O(\log(n))$, meaning that logarithmically sized
shingles should avoid communicationally expensive merges.

Thus, when the two strings are composed of random iid bits, then, under the appropriate choice
of $l$ from~\eqref{eq:shingleSize}, we can expect that no merging
is needed giving an overall communication complexity that is $O \left(\alpha \log^2(n) \right)$, for large $n$.

\subsection{Rateless approach}
\label{sec:rateless}
Observe that Protocol~\ref{protocol:string} communicates two types of data:  (i) set reconciliation data from step 2, and (ii) merged shingle indices in step 5.

The set reconciliation data can be ratelessly streamed for reconciling strings with arbitrary
edit distance by using a simple modification of the
protocol in~\cite{MTZ00}.  Specifically, a characteristic polynomial
\[
\charPoly{S_\sigma}{Z} = (Z-s_1)(Z-s_2)(Z-s_3)\cdots(Z-s_{|S_\sigma|})
\]
of the shingles $s_i \in S_\sigma$ is computed and its evaluations at points in an
appropriately sized finite field are provided to the decoder, which similarly computes
evaluations of its own characteristic polynomial.  The rational function representing
the division of the two polynomials can be determined from any $\Delta$ sample points,
if the two shingle sets differ in at most $\Delta$ shingles (an additional $k$ verification
points can be added to probabilistic check the result).

The merged shingle indices, which can be determined independently of the reconciliation,
 can be encoded with any standard rateless
code~\cite{Tornado_ACM,Digital_Fountain,raptor}, and the two rateless streams can be
combined by considering them inputs to yet a third rateless encoding.


\section{Conclusion}
\label{sec:conclusion}
We have provided a linear-time algorithm for determining whether a given string is
uniquely decodable from its bigrams.  Our algorithm is online, in that it needs only
constant-time pre-processing, and streaming, in that results for one string can be
sub-linearly extended to a superstring.  We have also shown how this algorithm
can be incorporated into an existing protocol for string reconciliation, though the
space of applications potentially extends further to networking, cryptography,
and genetic engineering.

Several interesting open questions remain.  For one, it is natural to ask whether
the proposed online algorithm can be extended for testing the existence of $2$,
$3$, ... or $k$ decodings.  It is also interesting to provide sharper bounds
for the numbers of merged shingles in Protocol~\ref{protocol:string} under different
random string models, as this could help determine the correct choice for
initial shingling size $l$, in addition to tightening bounds on the communication
complexity of the protocol.



\bibliography{common}
\bibliographystyle{plain}

\end{document}